\documentclass[12pt,prb,aps,amssymbols]{revtex4}
\usepackage{amssymb}
\usepackage{bm}
\usepackage{amsbsy}
\usepackage{epsfig}
\usepackage{color}
\usepackage{amsmath}
\begin{document}

\date{\today}

\title{ \begin{center}
Intrinsic rotation drive by collisionless trapped electron mode turbulence
\end{center}}
\author{Lu Wang,$^{1*}$ Shuitao Peng,$^{1}$ and P. H. Diamond$^{2}$\\
$^1$State Key Laboratory of Advanced Electromagnetic Engineering and Technology, School of Electrical and Electronic Engineering, Huazhong University of Science and Technology, Wuhan 430074, China\\
$^2$Center for Momentum Transport and Flow Organization and\\Center for Astrophysics and Space Sciences,\\ University of California at San Diego, La Jolla, CA 92093-0424, USA\\
$^*$E-mail: luwang@hust.edu.cn}
\maketitle \maketitle

\section*{Abstract}


Both the parallel residual stress and parallel turbulent acceleration driven by electrostatic collisionsless trapped electron mode (CTEM) turbulence are calculated analytically using gyrokinetic theory. Quasilinear results show that the parallel residual stress contributes an outward flux of co-current rotation for normal magnetic shear and turbulence intensity profile increasing outward. This may induce intrinsic counter-current rotation or flattening of the co-current rotation profile. The parallel turbulent acceleration driven by CTEM turbulence vanishes, due to the absence of a phase shift between density fluctuation and ion pressure fluctuation. This is different from the case of ion temperature gradient (ITG) turbulence, for which the turbulent acceleration can provide co-current drive for normal magnetic shear and turbulence intensity profile increasing outward. Its order of magnitude is predicted to be the same as that of the divergence of the residual stress [Lu Wang and P.H. Diamond, Phys. Rev. Lett. {\bf 110}, 265006 (2013)]. A possible connection of these theoretical results to experimental observations of electron cyclotron heating effects on toroidal rotation is discussed.

\section{Introduction}

Intrinsic (or, spontaneous) plasma rotation is of great interest in magnetic confinement fusion. Plasma rotation is thought to play a critical role in both stabilizing macroscopic magnetohydrodynamic (MHD) instabilities, such as resistive wall modes\cite{Bondeson,Betti} and in reducing turbulent transport level due to plasma microturbulence and thereby improving the performance of plasma confinement. Intrinsic rotation is particularly important for the International Thermonuclear Experimental Reactor, because conventional neutral beam injection (NBI) may not provide sufficient rotation drive due to its limited beam penetration. Therefore, understanding the underlying physical mechanisms for intrinsic rotation generation is an outstanding issue and there has been intensive research in intrinsic rotation in the magnetic fusion energy community in recent years (see Refs.~\onlinecite{Diamond2013,Ida2014} for overviews).

Experimentally, electron cyclotron heating (ECH) effects on toroidal rotation have been studied in many tokamaks such as DIII-D\cite{deGrassie2004,deGrassie2007}, TCV\cite{Porte2007}, JT-60U\cite{Sakamoto2006,Yoshida2009}, AUG\cite{McDermott1, McDermott2}, KSTAR\cite{Shi2013,Shi2016} for various conditions, and in stellarators\cite{Ida2001,Ida2015} as well. The direction of core toroidal rotation changes from co-current in Ohmic H-mode plasmas to counter-current in ECH H-mode plasmas.\cite{deGrassie2004,deGrassie2007} It is also found that ECH induces an increment of counter-current rotation in co-current or balanced NBI heated L-mode and H-mode plasmas.\cite{Yoshida2009,McDermott1,McDermott2,Shi2013,Shi2016} After turning on ECH, the electron temperature (and hence its gradient) in the core region increase dramatically, and the density gradient also steepens, while the ion temperature and its gradient just decrease slightly. Counter-current torque induced by micro-turbulence rather than damping by MHD activity as a physical mechanism to explain ECH effects on toroidal rotation has been discussed in Refs.~\onlinecite{Ida2015, Shi2016}. Linear stability study by gyrokinetic simulation suggests a possible transition from ion temperature gradient (ITG) turbulence to trapped elctron mode (TEM) turbulence due to the changes in temperature and density profiles induced by ECH.\cite{McDermott2,Shi2013} It implies that the counter-current effect induced by ECH might be related to the turbulence transition from ITG to TEM. Stronger TEM excitation corresponding to larger reduction of co-current toroidal rotation due to larger trapped electron fraction for off-aix ECH case is reported in Ref.~\onlinecite{Shi2016}. Due to lack of direct fluctuation measurements corresponding to the turbulence mode transition, it is difficult to quantitatively compare experimental observations with theoretical and simulation results.

From theory and numerical simulations, it is widely recognized that tokamak intrinsic rotation can be self-generated by micro-turbulence. In turbulent momentum flux, the non-diffusive, non-convective component, which is usually referred as residual stress is thought to be the origin of intrinsic torque. Intrinsic rotation generation due to residual stress driven by electrostatic ITG turbulence has been studied both by gyrokinetic theory\cite{Kosuga2010} and by gyrokinetic simulation.\cite{WangWX2009,Ku2012} Collisionless trapped electron mode (CTEM) turbulence driven intrinsic torque associated with residual stress was also reported in Refs.~\onlinecite{Diamond2008,WangWX2010}. A local gyrokinetic study on the relationship between the changes of intrinsic rotation and the turbulence transition from ITG to TEM is also based on an explanation by residual stress.\cite{Angioni2011} In addition, turbulent acceleration in ITG turbulence was proposed as a new possible mechanism for driving intrinsic rotation.\cite{WangL2013,Garbet2013} In Ref.~\onlinecite{WangL2013}, turbulent acceleration for parallel flow velocity is obtained, which cannot be written as a divergence of stress term, therefore, it acts as a local source or sink. This is different from the physics of residual stress, which enters the flow velocity equation via its divergence. In other words, the turbulent acceleration is an effective volume-force, while the residual stress is a kind of surface stress. This bears some similarity with the difference between the turbulent heating\cite{Manheimer1977,Zhao2012} (a possible heat source) and the turbulent energy pinch\cite{Wang2011} (one component of the energy flux). Recently, turbulent acceleration for mean parallel flow is extended to electromagnetic turbulence.\cite{PengWang} One may wonder if the turbulent acceleration contradicts with momentum conservation. The answer is not. Basically, this is because the conserved physical quantity is canonical momentum density but not the flow velocity. The canonical momentum density conservation should be obtained by summing the canonical momentum equation over all species and using quasinetrality condition.\cite{Sugama1998,Parra2010,Scott2010,Brizard2011,Abiteboul2011} However, the mean parallel flow velocity evolution equation describes velocity rather than canonical momentum density. Therefore, there is no constraint of conservation for parallel flow velocity. The turbulent acceleration due to CTEM turbulence has not been calculated. Hence, the goal of this work is to study the theory of intrinsic rotation driven by both the residual stress and the turbulent acceleration in CTEM turbulence. A quantitative comparison with the experimental observations beyond the scope of this paper.

In this work, following the procedures of Ref.~\onlinecite{WangL2013}, a mean parallel flow velocity evolution equation is derived for electrostatic CTEM turbulence using gyrokinetic theory.
The principal results of this paper are as follows:
\begin{enumerate}
\item  Both parallel Reynolds stress and parallel turbulent acceleration are estimated using quasilinear theory. Parallel symmetry breaking induced by fluctuation intensity gradient\cite{Gurcan2010} is invoked to calculate the residual stress.
\item The parallel residual stress for CTEM turbulence is predicted to contribute an outward flux of co-current rotation for normal magnetic shear and turbulence intensity profile increasing outward. This outward flux of co-current rotation may lead to flattening of co-current rotation or intrinsic counter-current rotation.
\item The turbulent acceleration driven by CTEM turbulence vanishes, since there is no phase shift between density fluctuation and ion pressure fluctuation. This is different from ITG turbulence for which the turbulent acceleration is a co-current drive for normal magnetic shear and turbulence intensity profile increasing outward, and its order of magnitude can be comparable to that of the divergence of residual stress.\cite{WangL2013}
\item If the turbulence mode transitions from ITG to TEM, vanishing co-current drive from turbulent acceleration and the outward flux of co-current rotation may cause flattening of the co-current rotation profile or counter-current rotation increment, which is qualitatively consistent with experimental observations of ECH effects on toroidal rotation.
\end{enumerate}

The remainder of this paper is organized as follows. In Sec.~II, we
present the derivation of the mean parallel flow velocity evolution equation. In Sec.~III, quasilinear estimates for both the residual stress and the turbulent acceleration for CTEM turbulence are given. Finally, we summarize our
work and discuss its possible implications for experiments in Sec.~IV.

\section{Parallel flow velocity evolution equation}

In this work, we investigate the evolution equation of parallel flow velocity rather than that of parallel momentum density. We start from nonlinear electrostatic gyrokinetic equation in the continuity form\cite{Hahm88}
\begin{equation}\label{GC Vlasov}
\frac{\partial }{\partial t} \left(F B^*\right)+
{\bf  \nabla}\cdot\left(\frac{d \bf  R}{dt} FB^*\right) +
\frac{\partial }{\partial  v_{\|}} \left(\frac{d  v_{\|}}{dt}F B^* \right) = 0,
\end{equation}
with gyrocenter equations of motion
\begin{equation}
\frac{d \bf  R}{dt} = v_{\|} \hat{\bf b} + \frac{c}{eB^*} \hat{\bf b} \times \left( e\nabla \langle\langle\delta \phi\rangle\rangle + \mu \nabla B + m_i v_{\|}^2 \hat{\bf b} \cdot \nabla \hat{\bf b} \right),
\end{equation}
and
\begin{equation}
\frac{d  v_{\|}}{dt} = - \frac{\bf B^*}{m_i B^*} \cdot \left(e \nabla \langle\langle\delta \phi\rangle\rangle + \mu \nabla B\right).
\end{equation}
Here, $F=F({\bf R}, \mu, v_{\|}, t)$ is the gyrocenter distribution function, $\mu$ is the gyrocenter magnetic moment, ${\bf B^*} = {\bf B} + \frac{m_ic}{e}v_{\|} \nabla \times \hat{\bf b}$, $B^* = \hat{\bf b} \cdot {\bf B^*}$ is the Jacobian of the transformation from the particle phase space to the gyrocenter phase space, and $\langle\langle\cdots\rangle\rangle$ denotes gyroaveraging.

By taking the zeroth order moment of the nonlinear gyrokinetic equation, we obtain the equation for gyrocenter density, $n\equiv\left(2\pi / m_i\right) \int d\mu d v_{\|} FB^*$,
\begin{equation}\label{density}
\frac{\partial}{\partial t} n + \nabla \cdot \left[\left(U_{\|} \hat{{\bf b}} + \delta {\bf v}_{E\times B} + {\bf v}_{d\kappa} + {\bf v}_{d\nabla} \right) n \right] =0.
\end{equation}
Then, we take the first order moment to obtain the equation for gyrocenter parallel momentum per ion mass,\cite{Hahm2007} $nU_{\|} \equiv \left(2\pi / m_i\right) \int d\mu d v_{\|} FB^* v_{\|}$,
\begin{eqnarray}\label{momentum}
&&\frac{\partial}{\partial t} \left(nU_{\|}\right) + \nabla \cdot \left[ \frac{P_i}{m_i}\hat{{\bf b}} + \left( \delta {\bf v}_{E\times B} + 3{\bf v}_{d\kappa} + {\bf v}_{d\nabla} \right) n U_{\|} \right] \nonumber\\
&=& - \left[\frac{e}{m_i} \hat{{\bf b}} \cdot \nabla \delta \phi + \frac{c}{B} \hat{{\bf b}} \times (\hat{{\bf b}} \cdot \nabla \hat{{\bf b}}) \cdot \nabla \delta \phi U_{\|} \right] n.
\end{eqnarray}
Here, $P_i= 2\pi \int d\mu d v_{\|} FB^* \left(v_{\|} - U_{\|}\right)^2 = \left(2\pi / m_i \right) \int d\mu d v_{\|} FB^* \mu B$ is the ion pressure, ${ \delta \bf v}_{E\times B} = c \hat{\bf b} \times \nabla \delta \phi /B$ is the fluctuating ${\bf E \times B}$ drift velocity, ${\bf v}_{d\kappa} =cT_i / \left(e B\right) \hat{{\bf b}} \times  \hat{{\bf b}} \cdot \nabla  \hat{{\bf b}}$ is the magnetic curvature drift velocity, ${\bf v}_{d\nabla} =cT_i / \left(e B^2\right)  \hat{{\bf b}} \times  \nabla B$ is the magnetic gradient drift velocity, and a long wavelength approximation $k_{\perp}^2 \rho_i^2 \ll 1$ has been used, with $k_{\perp}$ the perpendicular wave number, and $\rho_i$ the ion Larmor radius.  The terms on the right hand side represent the parallel electric force, along with the effective magnetic field ${\bf B}^*/B^*$. These terms cannot be written as a divergence of momentum flux, consistent with the interpretation as turbulent momentum source in Ref.~\onlinecite{Garbet2013}.

In this work, we focus on ion parallel flow velocity equation, but not gyrokinetic parallel momentum conservation\cite{Scott2010,Brizard2011,Abiteboul2011} or the ion parallel momentum equation.\cite{Garbet2013} The conserved quantity corresponding to toroidal symmetry is the total toroidal canonical momentum density carried by particles (summing over all species) and fields but not the ion flow velocity. Therefore, the presence of turbulent source or sink in the mean ion flow velocity equation does not contradict the gyrokinetic toroidal momentum conservation.\cite{Scott2010,Brizard2011,Abiteboul2011}  Although the most natural quantity for theoretical study is the toroidal canonical momentum density, the quantity measured and estimated from experimental observation is the toroidal ion flow velocity. The magnitude of toroidal flow velocity can be approximated by parallel flow velocity for tokamaks, since the toroidal field is much larger than the poloidal field. So taking $\left((\ref{momentum})- U_{\|} \times (\ref{density}) \right) / n$, we can obtain the ion parallel flow velocity equation\cite{WangL2013}
\begin{eqnarray}\label{velocity}
&&\frac{\partial}{\partial t} U_{\|} + \nabla \cdot \left[ \left(\delta {\bf v}_{E\times B} + 4{\bf v}_{d\nabla} \right)U_{\|} \right] \nonumber\\
&=& -\left[2{\bf v}_{d\nabla} \cdot \frac{\nabla n}{n} - \frac{e}{T_{i}} {\bf v}_{d\nabla} \cdot \nabla \delta \phi - 2 {\bf v}_{d\nabla} \cdot \frac{\nabla T_{i}}{T_{i}} \right] U_{\|}\nonumber\\
 &&- \frac{1}{m_i} \hat{{\bf b}} \cdot \left( e \nabla \delta \phi + \frac{1}{n} \nabla P_i\right).
\end{eqnarray}
In low-$\beta$ plasmas, $\hat{{\bf b}} \times  \left( \hat{{\bf b}} \cdot \nabla  \hat{{\bf b}} \right) \simeq (1/B)\hat{{\bf b}} \times  \nabla B$, so the magnetic curvature drift can be approximated as the magnetic gradient drift, i.e., ${\bf v}_{d\kappa}\simeq {\bf v}_{d\nabla}$. Note that the drift velocities are compressible in toroidal geometry. $\nabla \cdot {\delta \bf v}_{E\times B} \simeq 2 (e/T_{i}) {\bf v}_{d\nabla} \cdot \nabla \delta \phi$ and $\nabla \cdot {\bf v}_{d\nabla} = {\bf v}_{d\nabla} \cdot (\nabla T_i)/T_{i}$ are used when deriving the preceding equation. Then, the mean parallel velocity equation can be derived by taking a flux surface average of Eq.~(\ref{velocity}), i.e.,
\begin{equation}\label{mean velocity}
\frac{\partial}{\partial t} \langle U_{\|} \rangle + \nabla \cdot \Pi_{r,\|} = a_{\|},
\end{equation}
where $\Pi_{r,\|}$ is the total parallel Reynolds stress, and $a_{\|}$ is the parallel turbulent acceleration. The total parallel Reynolds stress can be written as
\begin{equation}\label{Reynolds total}
\Pi_{r,\|}=\left\langle \delta v_{E\times B,r}\delta U_{\|} \right\rangle + \left\langle 4 v_{d\nabla,r}^0\frac{\delta T_i}{T_{i0}} \delta U_{\|}\right\rangle.
\end{equation}
The two terms on the right hand side of Eq.~(\ref{Reynolds total}) come from the radial components of the fluctuating ${\delta \bf E\times B}$ velocity and the magnetic drift velocity, respectively. The second term was shown to be subdominant to the first one by Hahm et al.\cite{Hahm2007}, so we only keep the fluctuating ${ \delta \bf E\times B}$ induced parallel Reynolds stress in this work. The parallel turbulent acceleration can be written as
\begin{eqnarray}
a_{\|} &=& \frac{1}{m_i n_0^2} \left\langle \delta n \hat{\bf b} \cdot \nabla \delta P_i\right\rangle - 2\left \langle \frac{\delta T_i}{T_{i0}} {\bf v}_{d\nabla}^0 \cdot \nabla \frac{\delta n}{n_0}  \right\rangle \left \langle U_{\|}\right\rangle\nonumber\\
&& +\left\langle \delta U_{\|} {\bf v}_{d\nabla}^0 \cdot \nabla \left(\frac{e \delta \phi}{T_{i0}} - 2\frac{\delta n}{n_0} - 2\frac{\delta T_i}{T_{i0}} \right) \right\rangle.
\end{eqnarray}
This turbulent acceleration term \emph{cannot} be written as a divergence of the parallel Reynolds stress! It plays the role of a local source/sink of parallel rotation, and so is significant for parallel rotation. In particular, the first term in the turbulent acceleration is related to parallel gradient of ion pressure fluctuation, but is independent of the parallel velocity. In our previous work,\cite{WangL2013} it was shown that the physics of turbulent acceleration is fundamentally different from that of the residual stress, but the order of magnitude of turbulent acceleration can be comparable to that of divergence of the residual stress in electrostatic ITG turbulence. Although CTEM is driven by the trapped electron precession drift resonance,\cite{Adam1976} toroidal effects on rotation are not the foci of this work. The magnetic drift induced Reynolds stress was shown to be subdominant in Ref.~\onlinecite{Hahm2007}. The intrinsic turbulent acceleration induced by parallel ion pressure gradient is robust. It is present in whether the magnetic geometry is cylindrical or toroidal. Therefore, in the following, we only consider fluctuating ${\delta \bf E\times B}$ induced parallel Reynolds stress, $\left\langle \delta v_{E\times B,r}\delta U_{\|} \right\rangle$, and the parallel ion pressure gradient induced turbulent acceleration, $\frac{1}{m_i n_0^2} \left\langle \delta n \hat{\bf b} \cdot \nabla \delta P_i\right\rangle$.


\section{Quasilinear expressions for residual stress and turbulent acceleration}

In our previous work, we calculated the turbulent acceleration term in ITG turbulence.\cite{WangL2013} In this work, we calculate the quasilinear expressions for both the residual stress and the turbulent acceleration term in CTEM turbulence. The linearized perturbed ion distribution function in Fourier space can be written as
\begin{equation}\label{fper}
\delta f_{ik} = -i \frac{k_{\|} v_{thi} x_{\|}  - \omega_{*i}[1+\eta_i(x_{\|}^2 + x_{\perp}^2 - 3/2) - 2 \frac{L_n}{v_{thi}}\frac{\partial U_0}{\partial r}x_{\|}]}{-i[\omega_k - k_{\|}v_{\|} ]} \tau \delta \hat{\phi}_k F_{i0},
\end{equation}
where $x_{\|} = \left( v_{\|} - U_0 \right) / v_{thi}$ with $v_{thi}=\sqrt{2T_i/m_i}$, $x_{\perp} = \sqrt{\mu B/T_{i0}}$, $\eta_i = L_n/L_{T_i} $, $L_n ^{-1}= -\frac{\partial}{\partial r} \ln n_0$ is the density gradient scale length, $L_{T_i}^{-1}=-\frac{\partial}{\partial r} \ln T_i$ is the ion temperature gradient scale length, $\omega_{*i}= -k_{\theta}\frac{cT_i}{eBL_n}$ is the ion diamagnetic drift frequency, $\tau=T_e/T_i$ is the ratio of the electron temperature to the ion temperature, $\delta \hat{\phi}_k = \frac{e\delta\phi_k}{T_e}$, and $F_{i0}$ is assumed to be a shifted Maxwellian equilibrium distribution function as follows:
\begin{equation}
F_{i0}=n_0 \left(\frac{m_i}{2\pi T_i}\right)^{3/2}\exp\left(- x_{\|}^2 - x_{\perp}^2 \right).
\end{equation}
Since we do not consider the toroidal effects on parallel rotation, magnetic drifts have been neglected in Eq.~(\ref{fper}). For long wavelength modes, $k_{\theta} \rho_i < 1/q$, with $q$ being the safety factor, an approximation $|k_{\|}v_{\|}| > \omega_{di}$ can be justified. Here, $\omega_k=\omega_r + i \gamma_k$, with $\omega_r$ and $\gamma_k$ being the real frequency and linear growth rate of CTEM. In Ref.~\onlinecite{Adam1976}, the mode is sufficiently localized near rational surface, Landau resonance between waves and ions was neglected.
Therefore, taking the fluid ion limit, $|\omega_{k}| \gg |k_{\|} v_{\|}|$,  the perturbed ion distribution function can be simplified to:
\begin{equation}\label{fper2}
\delta f_{ik} = \frac{1}{\omega_{k}} \left\{k_{\|} v_{thi} x_{\|} + \frac{\omega_{*e}}{\tau} \left[1+\eta_i(x_{\|}^2 + x_{\perp}^2 - 3/2) - 2 \frac{L_n}{v_{thi}}\frac{\partial U_0}{\partial r}x_{\|}\right]\right\} \tau \delta \hat{\phi}_k F_{i0}.
\end{equation}

Now, we first calculate the parallel Reynolds stress, and determine its non-diffusive and non-convective component, i.e., residual stress. To obtain a quasilinear evaluation of the parallel Reynolds stress $\left\langle \delta v_{E\times B,r}\delta U_{\|} \right\rangle$, we need to calculate $\delta U_{\|}$. Taking the $\left( v_{\|}-U_0\right)$ moment of the fluctuation distribution yields
\begin{equation}
n_0 \delta U_{\|k} = n_0 k_{\theta} \rho_s c_s \frac{1}{\omega_{k}} \left(-\frac{\partial U_0}{\partial r} + \frac{k_{\|}}{k_{\theta}}\omega_{ci} \right) \delta \hat{\phi}_k .
\end{equation}
Then, we can obtain the Reynolds stress as follows:
\begin{eqnarray}\label{Reynolds stress}
\left\langle \delta v_{E\times B,r}\delta U_{\|}\right\rangle&=&  \Re \sum_k i k_{\theta}^2 \rho_s^2 c_s^2 \frac{1}{\omega_{k}} \left(-\frac{\partial U_0}{\partial r} + \frac{k_{\|}}{k_{\theta}}\omega_{ci} \right) \left|\delta \hat{\phi}_k\right|^2 \nonumber\\
&=& -\chi_{\phi} \frac{\partial U_0}{\partial r} + \Pi_{r,\|}^{res}.
\end{eqnarray}
Here, the first term is the diffusive term, with diffusivity
\begin{equation}
\chi_{\phi}= \sum_k \frac{|\gamma_k|}{\omega_{r}^2} k_{\theta}^2 \rho_s^2 c_s^2 I_k,
\end{equation}
and the second term is an off-diagonal term, which is the so called residual stress
\begin{equation}
\Pi_{r,\|}^{res}= \sum_k  \frac{|\gamma_k|}{\omega_{r}^2} k_{\|}k_{\theta} \rho_s c_s^3 I_k.
\end{equation}
Here, $I_k=\left|\delta \hat{\phi}_k\right|^2$ is the turbulence intensity, $\Re \left[\frac{i}{\omega_k}\right]=\frac{|\gamma_k|}{\omega_{r}^2}$ is used, and the absolute value of $|\gamma_k|$ is required by causality. There is no pinch term since toroidal effects have been ignored in the perturbed ion distribution function. It is known that the residual stress usually vanishes if the turbulence intensity $I_k$ is symmetric with respect to $k_{\|}$. The mechanism for $k_{\|}$ symmetry breaking has been intensively studied in the past few years. The mechanism includes: asymmetric instability,\cite{Stringer1964} $E\times B$ shear,\cite{Dominguez1993,Gurcan2007} polarization drift,\cite{McDevitt2009} up-down asymmetry of flux surfaces,\cite{Camenen2009,Peeters2005,Sugama2011,Parra2011} and turbulence intensity gradient.\cite{Gurcan2010} $E\times B$ shear is a frequently invoked symmetry breaking mechanism for the case of  edge transport barriers or internal transport barriers, due to their steep ion pressure profiles. However, the \emph{ion} pressure profile in CTEM regime need not to be very steep. Another simple symmetry breaking mechanism is that due to the turbulence intensity gradient.\cite{Gurcan2010} In this work, we focus the turbulent intensity gradient driven parallel residual stress for CTEM turbulence.
In toroidal geometry, $k_{\theta} = nq/r$ and $k_{\|} = k_{\theta} x \hat{s} / (q R_0)$, where $\hat{s}$ is the magnetic shear, $x=r_{m,n} - r$, and $r_{m,n}$ is the radial location of the resonant surface. Proceeding as in the study of the residual stress induced by intensity gradient,\cite{Gurcan2010} i.e., $I_k(x)=|\phi_k|^2(x) = I_k(0) + x (\partial I_k / \partial x)$, it follows that the residual stress can be written as
\begin{eqnarray}
\Pi_{r,\|}^{res} &=& \sum_k  \frac{|\gamma_k|}{\omega_{r}^2} k_{\theta}^2 \rho_s c_s^3 \frac{\hat{s}}{qR_0} x^2 \frac{\partial I_k}{\partial x}.
\end{eqnarray}
We note that the residual stress driven by CTEM turbulence is an outward flux of co-current rotation for normal magnetic shear and $\partial I_k / \partial x >0$.  This may result in flattening of co-current rotation or an intrinsic counter-current rotation increment. Flattening effects due to residual stress are more important for stronger CTEM excitation. This is qualitatively consistent with experimental results.\cite{Shi2016}

Next, to calculate the turbulent acceleration induced by parallel gradient of ion pressure fluctuation, $a_{\|} =\frac{1}{m_i n_0^2} \left\langle \delta n \hat{\bf b} \cdot \nabla \delta P_i\right\rangle $, we need to obtain $\delta P_i$ by taking the second order moment of the perturbed ion distribution function, i.e., Eq.~(10),
\begin{eqnarray}
\delta P_{ik}&=&\frac{1}{3}  \int d^3v \left[ m_i \left(v_{\|} - U_0\right)^2 + 2 \mu B \right] \delta f_{ik},\nonumber\\
&\simeq& P_{i0} \frac{\omega_{*e}}{\omega_{r}}\left(1- i \frac{|\gamma_k|}{\omega_r} \right) \left(1+ \eta_i\right) \delta \hat{\phi}_k .
\end{eqnarray}
The ion gyrocenter density fluctuation can be obtained from quasi-neutrality condition without consideration of finite Larmor radius (FLR) effects, i.e.,
\begin{equation}
\delta n_k = n_0 \left( 1- i \delta_k \right) \delta \hat{\phi}_k,
\end{equation}
where the first term on the right hand side is the adiabatic electron response, the second one is the non-adiabatic response with $\delta_k \simeq \frac{|\gamma_k|}{\omega_r}$.\cite{Horton1999} From Ref.~\onlinecite{Adam1976}, for long wavelength version ($k_{\theta} \rho_s \ll 1$) of CTEM which is mainly driven by electron temperature gradient, the real frequency can be approximated as $\omega_r \approx \omega_{*e}  =k_{\theta} \rho_s c_s / L_n$, with $L_n$ being the density gradient scale length, and the linear growth rate is
\begin{equation}
\left|\frac{\gamma_{k}}{\omega_{*e}}\right| = 2 \sqrt{\pi \epsilon} \left(\frac{R_0}{L_n G}\right)^{3/2} \exp \left(-\frac{R_0}{L_n G}\right) \eta_e  \left(\frac{R_0}{L_n G} - \frac{3}{2}\right),
\end{equation}
where $\epsilon$ is the inverse aspect ratio, $G$ is a function of magnetic shear $\hat{s}$ and azimuthal angle $\theta$ of the turning point of a trapped electron,\cite{Kadomtsev1967} and $\eta_e = L_n / L_{T_e}$ with $L_{T_e}$ the electron temperature gradient scale length. It is noted that there is no phase shift between the density fluctuation and the ion pressure fluctuation, so the turbulent acceleration vanishes to the lowest order, i.e.,
\begin{equation}
a_{\|} = \frac{1}{m_i n_0^2} \left\langle \delta n \hat{\bf b} \cdot \nabla \delta P_i\right\rangle \simeq 0.
\end{equation}
This means that for CTEM turbulence, intrinsic rotation drive from residual stress is necessarily dominant over that from turbulent acceleration. This is different from the ITG case where the turbulent acceleration can provide co-current intrinsic rotation drive, and its order of magnitude can be comparable to the divergence of residual stress. Unfortunately, since it is difficult for ion temperature fluctuation diagnostics, one cannot directly measure the turbulent acceleration from experiments.

\section{Summary and discussion}

In this work, we investigate the intrinsic parallel rotation generation by CTEM turbulence. By using the electrostatic gyrokinetic theory, we analytically derive the mean parallel flow velocity evolution equation which includes the usual parallel Reynolds stress and the parallel turbulent acceleration, as well. From quasilinear estimates, we find that the parallel residual stress for CTEM turbulence is an outward flux of co-current rotation for normal magnetic shear and positive fluctuation intensity gradient. The outward flux typically leads to an intrinsic rotation in the counter-current direction or flattening of co-current rotation. We found that the turbulent acceleration driven by CTEM turbulence vanishes, due to the absence of a phase shift between the density fluctuation and the ion pressure fluctuation. This is analogue to zero particle flux for adiabatic electrons for which there is no phase shift between density and electric potential fluctuations. Therefore, the turbulent intrinsic rotation drive in CTEM turbulence mainly comes from the residual stress but not the acceleration. The turbulent intrinsic rotation drive is sensitive to turbulence mode, which is also suggested by experimental observations. In this work, long wavelength limit is used, so the ion polarization density is neglected. Extension to a wide range of wavelength version of CTEM\cite{Hahm1991} by self-consistently taking into account finite Larmor radius effects on CTEM instability and turbulent intrinsic rotation drive may be worthwhile.

Finally, we discuss possible connections of our theoretical results to the experimental observations of ECH effects on core toroidal rotation. It was shown that the turbulent acceleration in ITG turbulence can provide co-current intrinsic rotation drive, and its order of magnitude is comparable to that of the divergence of residual stress.\cite{WangL2013} However, for CTEM turbulence, the turbulent acceleration vanishes, and the residual stress induces an outward flux of co-current rotation. Therefore, the turbulence mode transition from ITG to CTEM leads to reduction of co-current acceleration and an outward flux of co-current rotation. This may be relevant to the experimental observation of co-current rotation flattening induced by ECH.\cite{McDermott2,Shi2013,Shi2016} Our theoretical results is qualitatively consistent with experimental observations, while a quantitative comparison is beyond the scope this work.

\section*{Acknowledgments}

We are grateful to X. Garbet, Y. J. Shi, J. M. Kwon, and the participants in the Festival of Theory, Aix en Provence 2015 for fruitful discussions. This work was supported by the NSFC Grant No.~11305071, the MOST of China under Contract No.~2013GB112002, and U.S. DOE Grant No. DE-FG02-04ER54738.

\end{document}